\documentclass[manuscript,screen,sigconf]{acmart}
\usepackage{graphicx} % Required for inserting images
\usepackage{booktabs}
\usepackage{array}
\usepackage{multirow}
\usepackage{tabularx}
\usepackage{hyperref}

\title{Racial/Ethnic Categories in AI and Algorithmic Fairness: Why They Matter and What They Represent}

\author{Jennifer Mickel}
\email{jamickel@utexas.edu}
\affiliation{%
    \institution{University of Texas at Austin} \country{United States of America}
}
\setcopyright{acmlicensed}
\copyrightyear{2024}
\acmYear{2024}
\acmDOI{XXXXXXX.XXXXXXX}

\acmConference[ACM FAccT '24]{ACM Conference on Fairness, Accountability, and Transparency}{June 03--06,
  2024}{Rio de Janeiro, Brazil}

\begin{document}
\begin{abstract}
Racial diversity has become increasingly discussed within the AI and algorithmic fairness literature, yet little attention is focused on justifying the choices of racial categories and understanding how people are racialized into these chosen racial categories. Even less attention is given to how racial categories shift and how the racialization process changes depending on the context of a dataset or model. An unclear understanding of \textit{who} comprises the racial categories chosen and \textit{how} people are racialized into these categories can lead to varying interpretations of these categories. These varying interpretations can lead to harm when the understanding of racial categories and the racialization process is misaligned from the actual racialization process and racial categories used. Harm can also arise if the racialization process and racial categories used are irrelevant or do not exist in the context they are applied.

In this paper, we make two contributions. First, we demonstrate how racial categories with unclear assumptions and little justification can lead to varying datasets that poorly represent groups obfuscated or unrepresented by the given racial categories and models that perform poorly on these groups. Second, we develop a framework, CIRCSheets, for documenting the choices and assumptions in choosing racial categories and the process of racialization into these categories to facilitate transparency in understanding the processes and assumptions made by dataset or model developers when selecting or using these racial categories. 
\end{abstract}


\maketitle
\section{Introduction}
The utilization of racial and ethnic categories in the development of datasets and models facilitates the inclusion and documentation of diverse perspectives. Racial and ethnic categories are especially crucial for datasets and models in which race and ethnicity serve as relevant factors, may act as confounding variables, or enable the ability to audit for fairness using race and ethnicity for fairness purposes. For example, understanding the racial and/or ethnic target of hate speech is crucial for understanding the impact of hate speech, as hate speech can differ based on the race and/or ethnicity of the target \cite{nielsen2002subtle}. Similarly, in health, race is correlated with health outcomes \cite{barger2009relative}, and knowledge of a patient's race and ethnicity can help contextualize the patient's experience and health history \cite{okoro2021examining}. In algorithmic fairness settings, knowledge of an individual's race and ethnicity allows for auditing of existing datasets and systems, and many fairness toolkits, such as Fairlearn, rely on this data \cite{bird2020fairlearn, lee2021landscape}. Despite the benefit of race and ethnicity, little justification is provided for the racial and ethnic categories chosen and why these categories are most relevant for a dataset or model's particular domain. Furthermore, even if the choice of racial and ethnic categories is justified, even less discussion of how these racial and ethnic categories are assigned to individuals and what factors influence the racialization of people into these categories is given. Discussion of how people are assigned or racialized into these categories is crucial as the racialization of people into particular racial groups varies based on cultural context \cite{drnovvsek2021comparing}. Discussing this racialization process allows for understanding how the cultural context(s) and domain(s) affect people's placement and racialization into racial categories.

The racial and ethnic categorization schema used in datasets and models varies based on numerous factors. Some racial schemas used are binary, as in Black/non-Black, Black/White, and White/non-White, while others use multiple racial categories, as in Asian, Black, Hispanic, and White \cite{abdu2023empirical}. The racial and ethnic categories selected determine what racial and ethnic experiences are valued and will be traceable. In the binary setting, this often leads to the exclusion of people not racialized into these groups, and people with multiple racial identities are obscured. In the case of White/non-White, the experiences of non-White individuals are treated similarly since they are in the same category, even though it is evident that the experiences of non-White individuals vary drastically. For example, the experiences of Asians and Blacks within the US cultural context vary immensely \cite{chanbonpin2015between}. 

In this paper, we discuss in greater depth the effect of racial categorization choices on datasets and models, and we demonstrate the importance of documenting choices and motivations for racial categories by showcasing how ill-defined racial categories can affect datasets and model performance. 
% To address this, we develop CIRCSheets, a framework for documenting assumptions, choices, and motivations behind the selection of racial categories and the racialization of people into these racial categories.
Our work is motivated by previous scholarship on racial categories in algorithmic and AI fairness \cite{abdu2023empirical, benthall2019racial, hanna2020towards}  as well as by existing documentation frameworks for datasets and models \cite{bender2018data, chmielinski2022dataset, crisan2022interactive, diaz2022crowdworksheets, gebru2021datasheets, holland2020dataset, hutchinson2021towards,  mitchell2019model, pushkarna2022data}. We extend this work by focusing on how the choice of racial categorization and the racialization of people into the chosen racial categories affects how well-represented people are and, subsequently, dataset quality and model performance. To combat these effects, we develop CIRCSheets, a novel framework allowing developers of datasets or models to document their motivations behind why they selected certain racial categories and consider the effects of their choice in racial categories. 
\section{Background and Related Work} 
\subsection{Racial Categories: The Status Quo}
Racial categories used in datasets and models tend to align with the US cultural context \cite{abdu2023empirical, benthall2019racial, hanna2020towards}. Abdu et al. \cite{abdu2023empirical} identify two main choices for racial categorization: binary and more than two races. When binary racial categorization is chosen, it often operates under a Black/White axis. If the racial classification selected is more than two races, the racial categorizations tend to echo the US census \cite{abdu2023empirical, benthall2019racial}. The common categories used with multiple racial categories were Asian, Black, Hispanic, and White \cite{abdu2023empirical}.

The use of racial categories in datasets and models can help ensure a wide variety of perspectives are represented and considered. Furthermore, the presence of racial categories aids in analyzing, testing, and auditing datasets and models for disparities between racial groups. Without racial categories, these analyses along the axis of race would be challenging to conduct \cite{sandvig2014auditing, vecchione2021algorithmic}. Unfortunately, poorly defined racial categories can hinder actualizing these benefits \cite{nugraheni2021family, rebbeck2022distinct}. This can occur if a racial category comprises multiple groups whose experiences of racialization vary because the racial group no longer serves as a meaningful proxy for people's lived experiences within those groups.

An example of a racial category that comprises multiple groups who are racialized differently in the US cultural context is White. Individuals of Middle Eastern and North African (MENA) descent are categorized as White within the US despite many members of MENA not perceiving themselves to be White \cite{marks2020collecting}. Furthermore, within the US cultural context, their lived experience and racialization differ from people of European ancestry \cite{maghbouleh2022middle}. Having MENA as part of the White racial category obfuscates the experiences of members of MENA within datasets and models, preventing researchers from observing disparate health outcomes of this group \cite{awad2022lack}. Practitioners and researchers cannot see if a model performs poorly on MENA or if a dataset accounts for the experiences of people who are part of MENA. Most existing fairness toolkits require demographic information to audit algorithms, so practitioners who use these tools cannot audit their models for information on how the model performs on MENA \cite{lee2021landscape}.

A racial category can obfuscate people within that category when a multiracial ethnicity is treated as a racial category. For example, Latinx is a multiracial ethnicity, and the experiences of Latinxs can vary drastically based on the cultural context they are in and their race. For example, in the US, the experiences of lighter-skinned and darker-skinned Latinxs differ \cite{uzogara2021belongs}. Darker-skinned Latinxs racialized as Black in the US cultural context experience anti-Black discrimination from Latinxs and Whites \cite{hernandez2022racial}.
Placing all Latinxs into the Latinx category would obfuscate the experiences of darker-skinned Latinxs and prevent researchers and practitioners from observing whether datasets include darker-skinned Latinxs and if models perform poorly on darker-skinned Latinxs. 

\subsection{Race and Ethnicity}
Race and ethnicity, although similar, are two different concepts. Racial groups are differentiated by physical differences in certain social constructs \cite{bashi1997theory, omi2014racial}. Whereas ethnic groups, are differentiated based on social practices such as "language, religion, rituals, and other patterns of behavior" \cite[pp. 106]{bashi1997theory, omi2014racial, yu202027}. 
Often, ethnic categories are treated as racial categories, which can pose a problem when an ethnicity is not synonymous with a race, as in the case of panethnicities (defined in 
Section 3.3). 
For example, some Afro-Latinx individuals identify or are racialized as Latinx ethnically and Black racially \cite{hordge2020out}. This can lead to obfuscation for Afro-Latinxs and members of other multiracial panethnicities because it is unclear whether an individual's racial identity takes precedence over their ethnic identity or vice versa. 
 
Race and ethnicity, although they have no biological determinant, have real impacts on people's lives, ranging from their health to education to work \cite{mcchesney2015teaching, mukhopadhyay2013real, berger2015more, carter2017racial, wheeler2017racial}. Documenting race and ethnicity within datasets and models allows us to see how models perform on various races and ethnicities and helps audit the model for disparate impact. Furthermore, practitioners can train models using loss functions or other techniques that utilize race to help mitigate the oppression people of various racial and ethnic groups experience. Loss functions, used to train models, can be designed to help fulfill these goals \cite{keswani2021towards}.
Without knowledge of race and ethnicity, it is incredibly challenging to audit for disparate performance along the axes of race and ethnicity. 

\subsection{Racialization}
Racialization refers to the process by which racial meaning is given to people \cite{murji2005racialization}. Factors of physical difference, such as skin color and eye shape, among others, affect how people are racialized, as do accents \cite{clair2015sociology, omi2014racial, shuck2006racializing}. 
The process of racialization varies depending on cultural context, and relevant features in one context may be irrelevant in another \cite{telles2014black}. For example, the racial identification of Latinx adolescents and young adults shifts from adolescence to young adulthood and varies depending on generational time in the US, demonstrating that the process of racialization within the US and Latin American countries varies substantially enough for their racial identities to change \cite{irizarry2023race}. Furthermore, as time spent in the US increases, an individual's racial identity is less likely to shift \cite{irizarry2023race}.

Self-racial identification and external racialization differ. For example, the responses of Puerto Ricans and Dominicans to the race question on the 2010 US Census differ drastically, with respondents interpreting the question of race differently and using different aspects of race to answer the question \cite{roth2010racial}. This leads to racial self-identification that differs from how Puerto Ricans and Dominicans would be racialized based on their phenotype within the US \cite{roth2010racial}. This is due, in part, to different cultural contexts between the US, Puerto Rico, and the Dominican Republic \cite{roth2010racial}. For example, in the Dominican Republic, Black is used to describe Haitians \cite{itzigsohn2005immigrant}.
This leads to the racial self-identification of Hispanics on the US Census racialized as Black in the US cultural context to be a poor proxy for their physical features \cite{telles2018latinos}.

Salient features of racialization can differ based on the cultural context one is in. In the US, skin color plays a large role in racializing people into racial categories \cite{monk2021beholding, omi2014racial}. In Latin America, physical features other than skin color, such as hair texture and facial structure, play a part in racializing someone as Black, causing Latinx individuals with similar skin tones to be racialized differently due to other physical features such as hair texture and facial features \cite{hernandez2022racial}. Utilizing racial categories without discussing how people are racialized prevents us from understanding who comprises these racial categories and what factors affect whether people are racialized into particular categories and can lead to harm if we transpose different understandings of racial categories and racialization. 

\subsection{Racial Categories: Contextual Relevance and History}
The choice of racial categories in datasets and models is influenced by an array of sociotechnical factors, ranging from technical factors, such as model limitations, to contextual relevance, such as cultural context \cite{abdu2023empirical}. Datasets and models developed within the US cultural context tend to utilize racial groups relevant to the US cultural context but provide little justification for these choices \cite{abdu2023empirical}. Sometimes the US census is used as justification, as in Andrus et al. \cite{andrus2022demographic} or prior work, as in Yang et al. \cite{yang2020towards}, but most position cultural context as a sufficient justification of racial categories, as in Borradaile et al. \cite{borradaile2020whose}. 

Race has been central to political life in the United States \cite{omi2014racial}. This is evident through political discourse, legal history, and the US Census \cite{omi2014racial}. The census has been used as a tool to encode these values \cite{abdu2023empirical}, which is evident when observing the history of racial categories within the US Census. As an example, the Census of 1890 had four categories to classify people with African ancestry out of a total of eight categories \cite{schor2017counting}. This preoccupation with blackness in 1890 reflects the political climate within the southern states at the time \cite{lee1993racial}. The US Census of 1960 also reflects the political climate of the time, as Hawaii became a state in 1959 and Hawaiian and part-Hawaiian were added as racial categories to the US Census for the first time \cite{lee1993racial, o2000irreconcilable}. Observing the racial categories in the census over the years showcases how race within the US cultural context has shifted. Before 1860, the racial categories the census included were along the axis of Black and White, but as Asian immigrants immigrated to the US, Asian racial categories were added \cite{hochschild2008racial}. 

Racialization for certain groups varies depending on the context and domain. For example, the racialization of Filipinos varies by context \cite{ocampo2016latinos}. Some Filipinos identify culturally as Latinx rather than Asian, but within educational contexts, they tend to be treated as Asian rather than Latinx \cite{ocampo2016latinos, ocampo2013really}. 
This is seen in the literature for some studies racialize Filipinos as Asian, as in Baluran et al. \cite{baluran2023life} and Irizarry et al. \cite{irizarry2015utilizing} while others racialize Filipinos as Hispanic, as in Trevi{\~n}o \cite{trevino1987standardized}.

In addition to the context associated with the domain one operates in, racialization is affected based on the cultural context \cite{drnovvsek2021comparing}. For example, the experience of Central-East European immigrants differs between the UK and Japan \cite{drnovvsek2021comparing}. 
In addition to this, the experiences of certain groups within a racial category vary. For example, East Asians and South Asians are both racialized as Asian, yet their experiences differ, which leads Americans of Chinese descent to have a higher life expectancy than Americans of South Indian descent \cite{baluran2023life}.

Racial categories also differ based on country. Farquharson \cite{farquharson2007racial} discusses the racial formation of racial categories in the US, South Africa, and Australia, all of which are settler colonial states and identify race along a Black/White axis. Despite this, within each cultural context, people are racialized into the Black category differently. In South Africa, people of African ancestry who are mixed are racialized as colored, while in the US, they would be considered Black \cite{daya2013panel, khanna2010if}. In Australia, the Aboriginal peoples are racialized as Black, while in the US, they would not be \cite{farquharson2007racial}.
Lack of justification regarding racial categories prevents critical analysis of the sociological foundation of racial categories. 

\subsection{Researcher Justifications}
Abdu et al. \cite{abdu2023empirical} identify five existing categories of racial category justification in the algorithmic fairness literature. Researcher justifications fall under data availability, technical factors, appeals to prior scientific work, epistemic concerns, and contextual relevance \cite{abdu2023empirical}. 

The first two categories of justifications, data availability and technical factors, focus on limiting factors that affect racial category justification.
Data availability affects the racial categories researchers can choose because the choice of racial categories was made earlier during the data curation process. Furthermore, researchers and practitioners must rely on the information regarding racial categories and racialization provided with the data. In many cases, this means no information is provided \cite{abdu2023empirical}. Technical factors can affect the racial categories chosen because the model or algorithm may require or be limited to a certain number of features, as in the case of Friedler et al. \cite{friedler2019comparative} where their model required a binary racial category as the algorithm's sensitive attribute.

The last three categories of justification appeal to prior scientific work, epistemic concerns, and contextual relevance, focus on justifications related to the goal of the dataset or model and the domain(s) and cultural context(s) in which the dataset and model will be used. Appeals to prior scientific work utilize existing literature as justification for the racial categories used \cite{abdu2023empirical}. Justifications regarding epistemic concerns centering racial categories with greater scientific rigor, such as describing what features constitute a person's placement into a particular racial category \cite{abdu2023empirical}. Cultural context refers to the racial categories that are relevant in particular societies \cite{abdu2023empirical}. Oftentimes, there is an assumption of collective understanding that the racial categories chosen are salient for a certain cultural context. For example, datasets developed in the US cultural context, as in Borradaile et al. \cite{borradaile2020whose}, will justify their choice of racial categories by saying they are relevant to the US context.

\section{How racial/ethnic categories can affect datasets and models}
With the usage of racial and ethnic categories during dataset and model development, it is often unclear who fits into these categories due to the lack of discussion regarding assumptions about who is racialized into these categories. The cultural relevance and demographic makeup of these categories, as well as the multi-dimensionality of race and ethnicity, can impact a dataset's quality and a model's performance. Section 3.2 demonstrates how different demographic distributions, possible in broad or ill-specified racial and ethnic categories, can affect model accuracy on a group level.  
% Datasets that have racial/ethnic categories and models trained and/or tested using racial/ethnic cate
% Ambiguity in racial/ethnic groups can affect datasets and model performance. In this section, we demonstrate how unclear racial/ethnic categories can lead to differing performance based on the distribution of racial/ethnic groups in unclear categories. 
% In this section, we demonstrate how ambiguity in describing racial/
\subsection{The Effect of Cultural (Ir)relevance}
Cultural relevance is crucial when selecting racial categories. Racial categories vary depending on cultural context  \cite{farquharson2007racial}. If the racial categories selected for a cultural context are irrelevant to the domain(s) and context(s) they will be deployed in, the benefit of racial categories is lost, as racial categories lose their meaning when irrelevant.

Some racial categories, such as Black, may exist in multiple different cultural contexts, but the people placed into this category change depending on the context. A poorly defined definition of Black, which occurs when there is little to no discussion of how people are racialized into the category of Black, can lead to the usage of varying definitions of Black, especially if a dataset or model is used in a variety of cultural contexts. This has occurred in the US where people have been categorized as Black even though they would be racialized as white \cite{schor2017counting}.

To illustrate this effect, imagine a dataset or model is developed for the cultural contexts of the US, South Africa, and Australia, where Black is a culturally relevant racial category \cite{farquharson2007racial}. The developers are aware that Black as a racial category exists in each of these contexts and select the racialization process for the Black racial category to be culturally relevant to Australia, which refers to the Aboriginal people as Black \cite{farquharson2007racial}. The developers make this selection without conveying the racialization process of people into the Black category. Another group decides to use the dataset or model in the US or South Africa without understanding that people racialized into the Black category within this dataset or model are Aboriginal. This can lead to downstream issues or harm as the Black category is not relevant to the US or South African context since the racialization process differs from that of Australia. To prevent this from occurring, it is crucial to understand how people are racialized into each racial category of a dataset or model to understand if those racialization processes are culturally relevant to the domain(s) users of the dataset or model want to utilize it for. 

\subsection{The Effect of Distribution Shift in Broad Categories}
Abdu et al. \cite{abdu2023empirical} identify two main choices for racial categorization: binary and more than two races. Previous work using binary racial categorization utilizes Black/White, Black/non-Black, or white/non-White \cite{abdu2023empirical}. Non-Black and non-White are broad categories, and the possible sociodemographic distributions can vary drastically. With these racial categorization schemas, it becomes unclear which groups comprise non-Black and non-White. The non-White category could be comprised solely of Latinx individuals, or it could be comprised of both Black and Latinx individuals. Understanding the composition of broad racial categories and \textit{who} can be included in these categories is crucial. Otherwise, dataset quality and model performance metrics might differ if the distributions within these broad categories shift.

\begin{table*}
    \begin{tabularx}{\linewidth}{XX|XXXXXX|XX}
         {\bf Classifier} & {\bf Metric} & {\bf Asian} & {\bf African American} & {\bf Caucasian} & {\bf Hispanic} & {\bf Native American} & {\bf Other} & {\bf All} & {\bf Groups Trained On}\\
         \hline
         \multirow{5}{\linewidth}{{\bf Everyone}} & TPR (\%) & 100.0 & 73.6 & 50.6 & 42.2 & 100.0 & 42.9 & 62.9 & 62.9 \\
         & FPR (\%) & 0.0 & 39.7 & 15.2 & {\bf 22.8} & 0.0 & 24.4 & 27.8 & 27.8 \\
         & PPV (\%) & 100.0 & 66.8 & 70.9 & {\it 59.4} & 100.0 & {\it 54.5} & 67.0 & 67.0 \\
         & FDR (\%) & 0.0 & 33.2 & 29.1 & {\bf 40.6} & 0.0 & {\bf 45.5} & 33.0 & 33.0 \\
         & Acc (\%) & 100.0 & 67.2 & 70.3 & {\it 61.8} & 100.0 & {\it 62.3} & 67.8 & 67.8 \\
         \hline
         \multirow{5}{\linewidth}{{\bf Black/ White}} & TPR (\%) & 100.0 & {\bf 74.2} & {\bf 51.7} & 44.4 & 100.0 & 46.4 & {\bf 63.9} & 66.3 \\
         & FPR (\%) & 0.0 & 41.0 & 15.6 & {\bf 22.8} & 0.0 & 24.4 & 28.5 & 29.7 \\
         & PPV (\%) & 100.0 & 66.2 & 70.8 & 60.6 & 100.0 & 56.5 & 66.8 & 67.4 \\
         & FDR (\%) & 0.0 & 33.8 & 29.2 & 39.4 & 0.0 & 43.5 & 33.2 & 32.6 \\
         & Acc (\%) & 100.0 & 66.9 & {\bf 70.5} & 62.7 & 100.0 & 63.8 & 67.9 & {\bf 68.4} \\
         \hline
         \multirow{5}{\linewidth}{{\bf White/non-White (Hispanic + Other)}} & TPR (\%) & 100.0 & 71.8 & {\it 44.4} & {\it 35.6} & 100.0 & 42.9 & 59.5 & {\it 42.6} \\
         & FPR (\%) & 0.0 & 35.1 & 14.4 & {\it 15.8} & 0.0 & {\it 19.5} & 24.4 & 15.2 \\
         & PPV (\%) & 100.0 & {\bf 68.9} & 69.3 & {\bf 64.0} & 100.0 & {\bf 60.0} & 68.6 & 67.3 \\
         & FDR (\%) & 0.0 & {\it 31.1} & 30.7 & {\it 36.0} & 0.0 & {\it 40.0} & 31.4 & 32.7 \\
         & Acc (\%) & 100.0 & {\bf 68.5} & {\it 68.2} & 62.7 & 100.0 & 65.2 & {\bf 68.0} & 66.9 \\
         \hline
         \multirow{5}{\linewidth}{{\bf White/non-White (Hispanic)}} & TPR (\%) & 100.0 & {\it 69.7} & {\it 44.4} & {\it 35.6} & 100.0 & {\it 35.7} & {\it 57.9} & {\it 42.6} \\
         & FPR (\%) & 0.0 & {\it 34.8} & {\it 12.3} & {\it 15.8} & 0.0 & {\it 19.5} & {\it 23.5} & {\it 13.0} \\
         & PPV (\%) & 100.0 & 68.5 & {\bf 72.5} & {\bf 64.0} & 100.0 & 55.6 & {\bf 68.9} & {\bf 70.9} \\
         & FDR (\%) & 0.0 & 31.5 & {\it 27.5} & {\it 36.0} & 0.0 & 44.4 & {\it 31.1} & {\it 29.1} \\
          & Acc (\%) & 100.0 & 67.6 & 69.4 & 62.7 & 100.0 & {\it 62.3} & 67.7 & 68.1 \\
         \hline
         \multirow{5}{\linewidth}{{\bf Black/non-Black (Hispanic + Other)}} & TPR (\%) & 100.0 & 73.6 & 49.4 & {\bf 46.7} & 100.0 & {\bf 50.0} & 63.2 & 69.0 \\
         & FPR (\%) & 0.0 & 41.3 & 16.5 & 21.1 & 0.0 & 24.4 & 28.8 & 36.7 \\
         & PPV (\%) & 100.0 & 65.9 & 68.8 & 63.6 & 100.0 & 58.3 & 66.3 & {\it 65.3} \\
         & FDR (\%) & 0.0 & 34.1 & 31.2 & 36.4 & 0.0 & 41.7 & 33.7 & {\bf 34.7} \\
         & Acc (\%) & 100.0 & 66.5 & 69.1 & {\bf 64.7} & 100.0 & 65.2 & {\it 67.4} & {\it 66.1} \\
         \hline
         \multirow{5}{\linewidth}{{\bf Black/non-Black (Hispanic)}} & TPR (\%) & 100.0 & 73.9 & 50.0 & {\bf 46.7} & 100.0 & {\bf 50.0} & 63.6 & {\bf 70.7} \\
         & FPR (\%) & 0.0 & {\bf 41.6} & {\bf 16.9} & 21.1 & 0.0 & 24.4 & {\bf 29.1} & {\bf 38.4} \\
         & PPV (\%) & 100.0 & {\it 65.8} & {\it 68.5} & 63.6 & 100.0 & 58.3 & {\it 66.2} & 65.6 \\
         & FDR (\%) & 0.0 & {\bf 34.2} & {\bf 31.5} & 36.4 & 0.0 & 41.7 & {\bf 33.8} & 34.4 \\
         & Acc (\%) & 100.0 & 66.5 & 69.1 & {\bf 64.7} & 100.0 & 65.2 & {\it 67.4} & 66.2 \\
    \end{tabularx}
    \caption{
    Recidivism prediction performance is measured by the true positive rate (TPR), false positive rate (FPR), positive predictive value (PPV), false discovery rate (FDR), which is (1 - PPV), and accuracy (Acc). The groups in parentheses next to the category in "Classifiers" refer to the groups the logistic regression models are trained on. For example, Black/non-Black (Hispanic + Other) means that the non-Black category consisted of those in the categories of Hispanic or Other and the model was trained on Black, Hispanic, and Other data points. The {\bf bolded} numbers correspond to the classifier with the highest percentage for that particular metric, and the {\it italicized} numbers correspond to the classifier with the lowest percentage for that particular metric. If something occurs three or more times, it is not bolded or italicized even if it meets the criteria. Asian and Native American are the same for each classifier, so none of those metrics are bolded or italicized.}
    \label{tab:first}
\end{table*}
To demonstrate the impact of this, we use the dataset associated with COMPAS, an algorithm used to predict the recidivism risk of defendants, to train a logistic regression classifier using varying distributions of data based on the racial and ethnic categories in the dataset \cite{angwin2022machine}. Our logistic regression classifiers are trained race-blind and use a threshold of 0.5.
We test the logistic regression classifiers on each demographic group individually, all demographic groups, and the demographic groups trained on. Our results are showcased in \autoref{tab:first} which demonstrates that performance metrics vary based on the data each logistic regression model was trained on. The overall accuracy for all groups between the classifiers is within 1\%, but, per group, the difference between accuracies can range almost three times that for Hispanic and Other and two times that for African American and Caucasian. This means that the choice of racial categorization schema, racial categories, and who is racialized into these categories can have a real effect on whether someone is more likely to be correctly predicted to rescind. 
% These figures are echoed and intensified when observing the other performance metrics which is discussed in the \hyperref[first_appendix]{Appendix}. 
The true positive rate varies within 5\%, and the false positive rate varies within 7\% across all groups. These figures only increase when looking at each group individually. African American, Hispanic, and Other have higher false positive rates to begin with, so individuals in these groups would be more affected by this variation in false positive rates. The positive predictive value and false discovery rate vary by 2.6\% for all groups and up to almost double that for Hispanic (4.6\%) and Other (5.5\%).

This variation also occurs within racial categorization schemas. For White/non-White, the performance metrics can vary around 5\% when comparing Everyone, White/Black, White/non-White (Hispanic + Other), and White/non-White (Hispanic), which would all be valid distributions under the White/non-White categorization. Similar variation occurs for Black/non-Black when comparing Everyone, White/Black, Black/non-Black (Hispanic + Other), and Black/non-Black (Hispanic) which would all be valid distributions under the Black/non-Black categorization. Even in more specific racial and ethnic categories like Asian, Black, Latinx, Indigenous, Pacific Islander, and White this can transpire, for different distributions of various ethnic groups or racial groups can occur in these categories, which can also lead to variation in dataset quality and model performance. 

\subsection{The Effect of Racial Multi-Dimensionality and Panethnicity}
\label{section3}
Existing usage of racial categories in datasets and models rarely allows for multiracial and panethnic identities. Due to technical limitations \cite{abdu2023empirical}, each person is assigned a singular racial category and is rarely assigned more than one racial category. This leads multiracial individuals and their experiences to be obfuscated in either a racial category that comprises part of their racial experience or an 'Other' category where other multiracial individuals are placed, often with differing experiences of race \cite{ford2021monoracial, remedios2013finally}. This manifests in models as Wolfe et al. \cite{wolfe2022evidence} demonstrate that multiracial people are more likely to be assigned a racial or ethnic label of a minority group rather than a majority group.

Panethnic identities are, similarly, seldom adequately represented in the racial and ethnic categories used in datasets and models \cite{irizarry2023race}. Panethnicity refers to the identity that forms when different ethnic or tribal groups build institutions and identities across these ethnic groups' boundaries, leading to panethnicities comprised of people of various racial identities \cite{okamoto2014panethnicity}. There are numerous panethnicities, and Latinx is an example of a panethnicity \cite{mora2022identifies}. 

When panethnicities are included as a category in the chosen racial/ethnic categories selected or used by practitioners, the panethnic categories tend to be treated as a racial category regardless of the other racial identities members of panethnic groups may have. This leads the racial identities of members of this panethnicity to be unaccounted for and causes members of a panethnicity to be treated similarly due to their categorization, obfuscating the varying experiences of people that can be associated, in part, with their racial identity \cite{lopez2013killing}. This is readily seen within the US cultural context when Latinx as a category is used to solely represent the experiences of Latinx individuals, negatively affecting Afro-Latinxs, as their identities are obfuscated since often they are unable to select a racial category that best describes their racial identity and experience. Many Afro-Latinxs are not accepted as Latinx by their lighter-skinned peers, leading some Afro-Latinx individuals to find solidarity in Black communities where they feel more accepted \cite{hordge2020out}. Placing Afro-Latinxs solely in the Latinx category would prevent datasets and models from being able to account for these experiences of Afro-Latinxs.

\section{CIRCSheets: a documentation framework for Considerations in Racial Categorization Selection}
We present CIRCSheets, a framework to articulate the choices of racial categories to better position and understand the effect of racial categorization choices made in developing a dataset or model. This framework allows for an improved understanding of the assumptions and choices made by the users and developers of datasets and models, helping future dataset and model users understand whether the racial categories are relevant to their use case. Furthermore, our framework helps facilitate understanding surrounding the aspects of racialization and cultural context(s) considered when making choices of racial categorization and the previous assumptions made. This documentation allows for an improved understanding of the effect of the racial categories chosen and their racialization process while decreasing the likelihood of misaligned interpretations of the racial categorizations and the racialization processes from the creators of the dataset or model.

% Our framework helps position
\subsection{Categories}

\quad \textit{Considerations}
\begin{itemize}
    \item Consider how data availability and technical implementation affect how race and ethnicity can be represented in the dataset and/or model.
    \item Consider the domain(s) for which the dataset or model is developed for and how this affects the racial categories salient to these domain(s) and the racialization process(es).
    \item Consider how well the chosen racial categories represent the population(s) represented by the dataset or the population(s) affected by the model.
\end{itemize}
\quad \textit{Documentation Questions}
\begin{enumerate}
    \item What are the racial categories utilized?
    \item What is the motivation behind using these racial categories?
    \item Are multiracial ethnic categories utilized? 
    \item If multiracial ethnic categories are used, what is the motivation behind using these categories, and are they being treated as racial categories?
    \item Are people who select multiple racial categories considered multiracial? Are people who select one or more ethnic categories and one racial category considered multiracial?
    \item If so, what category are they placed into, and are other people who select multiple different racial and/or ethnic categories also placed into that same category? If not, what category are they placed in, and does ethnicity take priority over race?
    \item For models, what is the technical implementation of the racial and/or ethnic categories?
    \item How do ethnic groups fit into these racial categories?
    \item Can people be obfuscated by these racial categories? If so, do these groups experience erasure and is the model or dataset likely to interact with them?
\end{enumerate}
\subsection{Racialization}
\quad \textit{Considerations}
\begin{itemize}
    \item Consider what contexts the dataset or model will be used in and how this affects racialization.
    \item Consider what factors will be used in the racialization process and who determines an individual's racial identity.
    \item Consider what the most relevant factors of racialization are within the context(s) the dataset or model operates within.
\end{itemize}
\quad \textit{Documentation questions}
\begin{enumerate}
    \item Who determines an individual's racial categorization? Is it the individual? 
    \item Are physical characteristics asked of an individual?
    \item Is cultural background asked of an individual?
    \item In what ways could the existing racial information be partial or incorrect? What impact could this have on the dataset or model?
    \item If using an existing dataset and no racialization information exists, what was the source of the dataset, what cultural context was it developed in, and is there any existing scholarship on the racialization choices of that dataset?
\end{enumerate}
\subsection{Cultural Context}
\quad \textit{Considerations}
\begin{itemize}
    \item Consider how racial identification can change in the chosen cultural context(s) of your dataset or model. 
    \item Within the cultural context(s) the dataset or model operates in, consider what groups experience marginalization and how the choice of racial categories can affect what groups have visibility in the dataset or model.
    \item For data collection and dataset development, consider what viewpoints associated with racial identification you want to be represented within your dataset.
\end{itemize}
\quad \textit{Documentation questions}
\begin{enumerate}
    \item What cultural context(s) is this dataset or model developed for?
    \item Will this dataset or model be used in different cultural context(s)?
    \item If the dataset or model is used in different cultural context(s) or domains, is there any misrepresentation that can occur due to changes in racialization or racial categories within these different cultural contexts and domains?
\end{enumerate}
\subsection{Multi-racial and pan-ethnicity}
\quad \textit{Considerations}
\begin{itemize}
    \item Consider how multiracial individuals and multiracial panethnicities are represented within racial categories and whether the representation of these ethnicities can lead to obfuscation between people of different races within those panethnicities. 
    \item Consider representing racial categories and ethnicities separately.
    \item Consider the representation of multiracial individuals within the dataset or model and whether this reflects their lived experiences within society. 
    \item Consider whether technical limitations influence whether multiracial individuals can be adequately represented within models. 
\end{itemize}
\quad \textit{Documentation questions}
\begin{enumerate}
    \item How are multiracial individuals and multiracial panethnicities categorized within the dataset or model?
    \item Can more than one racial or ethnic category be selected?
    \item Do the categories given to panethnic individuals effectively communicate their racial and ethnic identities?
    \item Are there any individuals, such as Afro-Latinxs, who may be \textit{in}adequately represented by the racial categorizations chosen?
\end{enumerate}

\subsection{Knowledge and Positionality}
\quad \textit{Considerations}
\begin{itemize}
    \item Consider consulting community members and stakeholders about what racial categories best represent them and how erasure can manifest with fewer racial categories.
    \item Consider the epistemic goal of the dataset or model and how choices in racial categories contribute to this goal.
    \item Consider how the lived experiences of the dataset or model developers and researchers contribute to which racial categories are chosen.
    \item When developing a dataset, consider what racial categories of annotators and examples in the dataset are relevant.
\end{itemize}

\quad \textit{Documentation questions}
\begin{enumerate}
    \item What are the cultural backgrounds and cultural knowledge of the dataset or model developers? How familiar and/or knowledgeable are they with the cultural context(s) of the dataset or model they are developing?
    \item If CIRCSheets is completed by people other than the original dataset or model developers, what are their cultural backgrounds? How familiar and/or knowledgeable are they with the dataset or model's cultural context(s)? 
    \item If annotators or crowd workers are used to develop a dataset or provide feedback to a model, what are their cultural backgrounds? How familiar and/or knowledgeable are they with the cultural context(s) of the instances they annotate?
    \item What stakeholders, community members, or other resources were consulted when selecting the racial categories?
\end{enumerate}

\section{Case Study}
To demonstrate CIRCSheets in action, we apply our framework to the dataset associated with COMPAS using existing knowledge available about these datasets \cite{angwin2022machine, kennedy2022introducing}.\\

\noindent\fbox{
    \parbox{\linewidth}{\centering
        \bf{Categories}
    }
}\\

\textbf{What are the racial categories utilized?}\\
African-American, Asian, Caucasian, Hispanic, Native American, and Other.\\

\textbf{What is the motivation behind using these racial categories?}\\
No motivation is provided, but these categories seem to be taken from the US Census \cite{hanna2020towards}.\\

\textbf{Are multiracial ethnic categories utilized?} \\
Yes, Hispanic, a multiracial ethnicity, is treated as a race.\\

\textbf{If multiracial ethnic categories are used, what is the motivation behind using these categories, and are they being treated as a racial category?}\\
No motivation is provided by the dataset developers.\\

\textbf{Are people who select multiple racial categories considered multiracial? Are people who select one or more ethnic categories and one racial category considered multiracial?}\\
It is unclear if people can select multiple racial and/or ethnic categories, but in the dataset, each instance is assigned one racial or ethnic category. It is unclear whether people who select Hispanic and another race are considered multiracial. This is not discussed by the dataset developers \cite{hanna2020towards}.\\

\textbf{If so, what category are they placed into, and are other people who select multiple different racial and/or ethnic categories also placed into that same category? If not, why category are they placed in, and does ethnicity take precedence over race?}\\
It is unclear what happens if people select multiple racial categories. It is possible that only one racial category is chosen for the individual from the ones they selected, or they are automatically placed into the "Other" category. The dataset developers do not discuss how multiracial individuals are categorized.\\

\textbf{For models, what is the technical implementation of the racial and/or ethnic categories?}\\
This is not applicable, as a model is not being used.\\

\textbf{How do ethnic groups fit into these racial categories?}\\
The dataset developers do not discuss this, but it seems that it follows the US Census with people who are descendants from the Black ethnicities of Africa are placed into the African-American category, people who are descendants of Asian ethnicities are placed into the Asian category, people who are descendants from European ethnicities are placed into the Caucasian category, and people with Native American ancestry are placed into the Native American category \cite{marks2020collecting}. It seems that people with ancestry in Hispanic countries are placed into the Hispanic category \cite(who are descendants), but it is not clear under what circumstances someone is placed into the Hispanic category rather than another racial category.\\

\textbf{Can people be obfuscated by these racial categories? If so, do these groups experience erasure, and is the model or dataset likely to interact with them?}\\
Yes. Because this dataset is centered in the US cultural context MENA (Middle Eastern and North African) individuals are most likely to be racialized as Caucasian. As discussed in Section 2.1, the experiences of MENA differ from the experiences of white people in the US \cite{maghbouleh2022middle}. Thus, it would not be possible to examine racial bias against MENA within COMPAS.
There is also no category for Pacific Islanders, so it seems that those who identify as Pacific Islander would be placed into the "Other" racial category, which would obfuscate the experiences of Pacific Islanders. 

"Other" has been used as a proxy for Latinx in the past \cite{rodriguez2000changing}. It is unclear whether most individuals placed in the "Other" category in COMPAS identify with a particular ethnicity as was the case in the 1990 and 2000 US Censuses \cite{rodriguez2000changing}. If "Other" can be used as a proxy for a particular ethnicity, this would further obfuscate groups, such as Pacific Islanders, who would be placed into the "Other" category. \\

\noindent\fbox{
    \parbox{\linewidth}{\centering
        \bf{Racialization}
    }
}\\

\textbf{Who determines an individual's racial categorization? Is it the individual?}\\
It is unclear since the dataset developers do not discuss how a person's racial identity is determined \cite{angwin2022machine, hanna2020towards}\\

\textbf{Are physical characteristics asked of an individual?}\\
The dataset does not document physical characteristics, although race and gender are recorded. It is unclear if physical characteristics were asked of individuals to racialize them into a particular racial category.\\

\textbf{Is cultural background asked of an individual?}\\
It is not documented in the dataset.\\

\textbf{In what ways could the existing racial information be partial or incorrect? What impact could this have on the dataset or model?}\\
It is possible that some people racialized into the "Hispanic" or "Other" category were incorrectly racialized and should have been placed into another category. It is possible the features used to racialize people into categories were irrelevant to this particular domain. 

This could impact the dataset because the dataset could have incorrect information, which would affect models trained on the dataset. These models may learn incorrect associations that, if deployed, would negatively impact the people affected by the model's decision. Furthermore, if the dataset is partially incorrect, auditing models would be more challenging since it would be unclear what information within the dataset is useful and what is irrelevant.\\

\textbf{If using an existing dataset and no racialization information exists, what was the source of the dataset, what cultural context was it developed in, and is there any existing scholarship on the racialization choices of that dataset?}\\
The dataset was developed in Broward County, Florida within the US cultural context. Existing scholarship on COMPAS discusses how "we don’t know why the data take on a particular racial schema, nor do we have information about how defendants are racially categorized" \cite[pp. 502]{hanna2020towards}. Hanna et al. \cite{hanna2020towards} discuss how the racial category an individual is placed into can change within a police department, so it is unclear how accurate the racial categories in COMPAS are for each individual even if the racial categorization schema were clearly communicated.\\

\noindent\fbox{
    \parbox{\linewidth}{\centering
        \bf{Cultural Context}
    }
}\\

\textbf{What cultural context(s) is this dataset or model developed for?}\\
This dataset was developed in the US cultural context because it was developed in Broward County, Florida \cite{hanna2020towards}.\\ 

\textbf{Will this dataset or model be used in different cultural context(s)?}\\
It is possible this data may be used in different cultural contexts, but it seems unlikely as the dataset was created using US police records.\\

\textbf{If the dataset or model is used in different cultural context(s) and/or domains, is there any misrepresentation that can occur due to changes in racialization and/or racial categories in different cultural contexts and domains?}\\
Misrepresentation can occur if the dataset is used in different cultural contexts, as the racial categories seem chosen with the US cultural context in mind. Furthermore, it is unclear how these racial categories were developed and what aspects of racialization were most important in deciding what racial group people were placed into. This can become a greater issue if this dataset were used in a different cultural context. Furthermore, laws change depending on the country (and, in some cases, cities), so in different cultural contexts, some people may not have been included in the dataset in the first place because their crime would not have been a crime in a different context\\

\noindent\fbox{
    \parbox{\linewidth}{\centering
        \bf{Multiracial and Panethnicity}
    }
}\\

\textbf{How are multiracial individuals and multiracial panethnicities categorized within the dataset or model?}\\
It is unclear how they are categorized within the dataset. It seems that only one category can be selected, so multiracial individuals may be placed in the "Other" racial category, or one of their racial identities may be chosen as their racial category. Either of these choices can have downstream impacts because the experiences of these multiracial individuals placed into these categories may differ from other individuals within this category.

The only multiracial panethnicity considered in the dataset is Hispanic, and it is treated as a race. It is unclear how Black and white Hispanics would be categorized. Any categorization schema based on the singular racial categories provided could obfuscate identities. If the Hispanic category supersedes the African-American or Caucasian category, then the experiences of Black Hispanics would be obfuscated. If race supersedes, then the experiences of both Black and white Hispanics would be obfuscated by the racial categories they have been placed in since their experiences differ from other Black and white individuals.\\

\textbf{Can more than one racial category be selected?}\\
No.\\

\textbf{Do the categories given to panethnic individuals effectively communicate their racial and ethnic identities?}\\
No, because only one category can be selected.\\

\textbf{Are there any individuals, such as Afro-Latinxs, who would not be adequately represented by the racial categorizations chosen?}\\
Yes, any multiracial individual or any individual who is racialized outside of their panethnicity, like Afro-Latinxs.\\

\noindent\fbox{
    \parbox{\linewidth}{\centering
        \bf{Knowledge and Positionality}
    }
}\\

\textbf{What are the cultural backgrounds and cultural knowledge of the dataset or model developers? How familiar and/or knowledgeable are they with the cultural context(s) of the dataset or model they are developing?}\\
This is unknown as no information was released from Broward County, Florida regarding this.\\

\textbf{If CIRCSheets is filled out by people other than the original dataset or model developers, what are their cultural backgrounds? How familiar and/or knowledgeable are they with the cultural context(s) of the dataset or model?}\\
The individual filling this out is a Russian-American woman who grew up in the US cultural context, so she is familiar with US racial structures.\\

\textbf{If annotators or crowd workers are used, what are their cultural backgrounds? How familiar and/or knowledgeable are they with the cultural context(s) of the instances they annotate?}\\
This is unknown as no information was released about this from Broward County, Florida.\\

\textbf{What stakeholders, community members, or other resources were consulted when deciding the racial categories?}\\
This is unknown as no information was released about this from Broward County, Florida.\\

% \subsection{The GAB Hate Speech Corpus}
% \noindent\fbox{
%     \parbox{\linewidth}{\centering
%         \bf{Categories}
%     }
% }
% \noindent\fbox{
%     \parbox{\linewidth}{\centering
%         \bf{Racialization}
%     }
% }
% \noindent\fbox{
%     \parbox{\linewidth}{\centering
%         \bf{Cultural Context}
%     }
% }
% \noindent\fbox{
%     \parbox{\linewidth}{\centering
%         \bf{Multi-racial and Pan-ethnicity}
%     }
% }
% \noindent\fbox{
%     \parbox{\linewidth}{\centering
%         \bf{Knowledge and Positionality}
%     }
% }
\section{Conclusion}
In this work, we discuss the importance of racial and ethnic categories and demonstrate the effect these choices can have on dataset quality and model performance with different interpretations of racial categories and racialization processes. Therefore, to facilitate understanding of the racial categories and racialization processes used, we develop CIRCSheets as a documentation tool for developers to communicate their assumptions, motivations, and racialization understanding, as well as, potential pitfalls. This documentation allows future users to better understand the racial and ethnic categories documented and how people are placed into these categories, assisting them in determining whether they can use this information in future tasks, such as auditing datasets and models or deploying models to consumers. Dataset and model users can also use CIRCSheets to communicate their own understanding of existing racial categories when information regarding the racial categories and racialization process in existing datasets and models is unclear or does not exist. 
% \section{Research Ethics and Social Impact}\section{Positionality}
% This work was written by a Russian-American
\bibliographystyle{plain}
\bibliography{bib}

% \appendix
% \input{sections/appendix}

\end{document}